\documentclass[a4paper,12pt]{article}

\usepackage{amsfonts}
\usepackage{amsmath}
\usepackage{latexsym}
\usepackage{graphics}
\usepackage[pdftex]{graphicx}

\providecommand{\eqref}[1]{Eq. (\ref{#1})}

\newcounter{note}

\newcommand{\second}{\textrm{s}}

\newcommand{\meter}{\textrm{m}}

\newcommand{\tesla}{\textrm{T}}
\newcommand{\micro}{\mu}
\newcommand{\milli}{\textrm{m}}

\newcommand{\rmd}{\textrm{d}}

\newcommand{\vect}[1]{\mathbf{#1}}

\newcommand{\signal}{S}
\newcommand{\ssmag}{M_0}
\newcommand{\gyromag}{\gamma}
\newcommand{\dwfactor}{b}

\newcommand{\grad}{\vect{G}}
\newcommand{\gmax}{G_{\max}}
\newcommand{\dgrad}{{\vect{G}_d}}
\newcommand{\ddir}{{\hat{\vect{G}}_d}}
\newcommand{\cgrad}{{\vect{G}_c}}
\newcommand{\sgrad}{{\vect{G}_s}}
\newcommand{\dlen}{\delta_d}
\newcommand{\clen}{\delta_c}
\newcommand{\slen}{\delta_s}
\newcommand{\gapone}{\tau_1}
\newcommand{\gaptwo}{\tau_2}
\newcommand{\tmix}{\tau_m}
\newcommand{\te}{{\tau_e}}
\newcommand{\tone}{T_1}
\newcommand{\ttwo}{T_2}
\newcommand{\diff}{d}

\newcommand{\tr}{{\tau_r}}
\newcommand{\tdd}{\tau_{dd}}
\newcommand{\tdc}{\tau_{dc}}
\newcommand{\tds}{\tau_{ds}}
\newcommand{\tcc}{\tau_{cc}}
\newcommand{\tcs}{\tau_{cs}}
\newcommand{\tss}{\tau_{ss}}
\newcommand{\pd}{\vect{e}_1}

\newcommand{\dirconc}{\eta}

\begin{document}

\bibliographystyle{unsrt}

\title{Diffusion imaging with stimulated echoes: signal models and experiment design}

\author{Daniel C. Alexander$^1$ \and Tim B. Dyrby$^2$ \\$^1$Centre for
Medical Image Computing,\\Department of Computer Science, University
College London,\\Gower Street, London WC1E 6BT, UK.\\$^2$Danish
Research Centre for Magnetic Resonance, Copenhagen\\University
Hospital Hvidovre, Hvidovre, Denmark.}

\date{}

\maketitle

Correspondence: D.Alexander@cs.ucl.ac.uk.

\clearpage

\begin{abstract}

Purpose: Stimulated echo acquisition mode (STEAM) diffusion MRI can be advantageous over pulsed-gradient spin-echo (PGSE) for diffusion times that are long compared to $\ttwo$. It is important therefore for biomedical diffusion imaging applications at 7T and above where $\ttwo$ is short. However, various gradient pulses in the STEAM sequence related to imaging contribute much greater diffusion weighting than in PGSE, but are often ignored during post-processing. We demonstrate here that this can severely bias parameter estimates.

Method: We present models for the STEAM signal for free and restricted diffusion that account for crusher and slice-select (butterfly) gradients to avoid such bias. The butterfly gradients also disrupt experiment design, typically by skewing gradient-vectors towards the slice direction. We propose a simple compensation to the diffusion gradient vector specified to the scanner that counterbalances the butterfly gradients to preserve the intended experiment design.

Results: High-field data from a fixed monkey brain experiments demonstrate the need for both the compensation during acquisition and correct modelling during post-processing for both diffusion tensor imaging and ActiveAx axon-diameter index mapping. Simulations support the results and indicate a similar need in in-vivo human applications.

Conclusion: Correct modelling and compensation are important for practical applications of STEAM diffusion MRI.

\end{abstract}


\clearpage

\section{Introduction}\label{sec:introduction}

Stimulated echo acquisition mode (STEAM) diffusion MRI~\cite{TannerJCP70,MerboldtMRM91} offers advantages over the more common pulsed-gradient spin-echo (PGSE) diffusion MRI when $\ttwo$ is short compared to the diffusion time and $\tone>\ttwo$.  Whereas $\ttwo$-decay occurs throughout the PGSE sequence, in STEAM the signal decays instead with rate $\tone$ during the mixing time $\tmix$, which determines the diffusion time.  Thus, despite a factor of two reduction in signal from loss of one coherence pathway, STEAM retains more signal than PGSE for large enough diffusion time.

STEAM diffusion MRI is common in tissue with short $\ttwo$, such as muscle or cartilage, e.g.~\cite{BrihuegaMorenoMRM03}. Current in-vivo human-brain diffusion MRI applications usually do not benefit from STEAM, because $\ttwo$ at 1.5T or 3T is relatively long compared to typical diffusion times. However, $\ttwo$ decreases and $\tone$ increases as field strength increases. Early evidence~\cite{DhitalISMRM10} already suggests benefits of STEAM for in-vivo human-brain diffusion tensor imaging (DTI) at 7T. Ex-vivo $q$-space studies of brain tissue, e.g.~\cite{StaniszMRM97,AssafMRM08,OngNIMG10}, usually prefer STEAM over PGSE, because they use high field preclinical scanners (i.e. research machines used for development rather than clinical practice), tissue fixation further reduces $\ttwo$, and lower sample temperature reduces diffusivity increasing necessary diffusion times~\cite{DyrbyHBM11,ShepherdMRM09}. Translation of advanced diffusion MRI techniques to high field in-vivo human applications is likely to rely on STEAM in place of PGSE. For example, diffusion spectrum imaging~\cite{WedeenNIMG08} requires high $\dwfactor$-values and consequently long echo time in PGSE, which becomes infeasible as $\ttwo$ decreases. Also, microstructure imaging techniques, such as ActiveAx~\cite{AlexanderNI10} and AxCaliber~\cite{AssafMRM08}, require long diffusion times to ensure sensitivity to large diameter axons~\cite{DyrbyMRM12}. 

Various gradients required for imaging in diffusion MRI add diffusion weighting and `cross-terms' in the $\dwfactor$-matrix~\cite{MattielloMRM97}. The most significant contributions are usually from the crusher and slice-select gradients, so-called `butterfly gradients'. In PGSE, their contribution to the diffusion weighting is usually negligible in practice, because the diffusion time for the butterfly gradients is only a few milliseconds (the length of the refocussing pulse).  However, in STEAM, that contribution is typically much more significant, because the diffusion time is approximately $\tmix$. Nevertheless, previous work with STEAM diffusion MRI, such as~\cite{AssafMRM08,OngNIMG10}, follows standard practice for PGSE and ignores the effect.

In this paper, we derive models for the STEAM signal that account for the diffusion weighting of the butterfly gradients and avoid unnecessary bias in parameter estimation during post-processing. Specifically, we adapt the DTI $\dwfactor$-matrix calculations in~\cite{MattielloMRM97} for STEAM and we derive new models for signals arising from restricted diffusion using the Gaussian phase distribution (GPD) approximation. The latter extend standard PGSE models for restricted diffusion in spheres~\cite{MurdayJCP68}, cylinders~\cite{VanGelderenJMR94}, and more general restricting geometries~\cite{StepisnikPhysicaB93} for use with STEAM. In addition, we propose a simple compensation of the diffusion gradient vector during acquisition that counterbalances the diffusion weighting of the butterfly gradients. This avoids disruptions to the experiment design (the intended set of $\dwfactor$-values, gradient directions, etc), which arise from the butterfly gradients skewing the effective diffusion weighting towards the slice direction.

Simulation and fixed-brain experiments use DTI and ActiveAx, orientationally invariant axon density and diameter index mapping, to demonstrate that ignoring the butterfly gradients in STEAM post processing biases parameter estimates. Using the new models avoids unnecessary bias. Moreover, these high angular resolution diffusion imaging (HARDI) applications demonstrate how the compensation corrects significant disruption to the experiment design leading to further improvements in accuracy and precision of parameter estimates. In combination, the new models and the compensation provide the essential tools for using STEAM in a wide range of practical applications. 

\section{Methods}\label{sec:methods}

This section introduces the STEAM diffusion-weighted pulse sequence. It then outlines various candidate signal models for both free, the diffusion tensor (DT) model, and restricted diffusion that support parameter estimation from measured data. The last subsection specifies the compensation for preserving experiment design.

\subsection{STEAM pulse sequence}\label{sec:sequence}

\begin{figure}
\includegraphics[width=\linewidth]{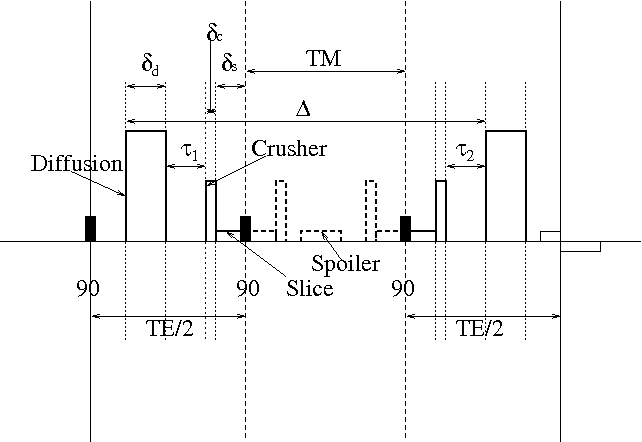}
\caption{\label{fig:STEAMSequence}Diagram of the STEAM pulse sequence.  See section~\ref{sec:sequence} for definitions and explanations.  TE is the echo time. TM is the mixing time, which is $\tmix$ in the main text.}
\end{figure}

The signal models in subsequent sections assume the idealized STEAM pulse sequence in figure~\ref{fig:STEAMSequence}, which consists of:

\begin{enumerate}

\item An initial $90^\circ$ pulse to tip the spins into the transverse plane.

\item A diffusion gradient pulse with duration $\dlen$ and constant gradient vector $\dgrad$, which starts at time zero.

\item A gap of length $\gapone$ with no gradients.

\item A crusher pulse, which starts at time $\dlen + \gapone$ and lasts for time $\clen$ with constant gradient vector $\cgrad$.

\item A slice-select pulse, which starts immediately after the crusher pulse at time $\dlen + \gapone + \clen$.  The slice-select pulse lasts for $2\slen$ with constant gradient vector $\sgrad$.  However, the second $90^\circ$ pulse occurs at the centre of the slice-select pulse, so only the first half contributes diffusion weighting.  Thus, to calculate the diffusion weighted signal, we consider the slice-select pulse to have length $\slen$, starting at $\dlen + \gapone + \clen$. 

\item A mixing time $\tmix$, which starts after the second $90^\circ$ pulse.

\item A spoiler pulse during the mixing time, which contributes no diffusion weighting so we do not consider it further.  Additional crusher pulses also occur during $\tmix$ that do not contribute diffusion weighting. Non-contributing pulses are dashed in figure~\ref{fig:STEAMSequence}.

\item A third $90^\circ$ pulse, which occurs at the end of the mixing time at the centre of a second slice-select pulse. The diffusion weighting part of the second slice-select is equal to the first, starting at time $\dlen + \gapone + \clen + \slen + \tmix$.

\item A second crusher gradient equal to the first at time $\dlen + \gapone + \clen + 2\slen + \tmix$.

\item A gap of length $\gaptwo$ with no gradients.

\item A second diffusion gradient pulse equal to the first at time $\dlen + \gapone + \gaptwo + 2\clen + 2\slen + \tmix$.

\end{enumerate}

\subsection{Signal models}\label{sec:signal}

We consider three approximations to the signal that account for the butterfly gradients in different ways:

\begin{itemize}

\item Approximation 1 (A1) ignores the butterfly gradients and considers only the diffusion gradients.

\item Approximation 2 (A2) identifies an effective diffusion gradient $\dgrad^\prime$ that incorporates the diffusion weighting of the diffusion and butterfly gradients. A simple choice is
\begin{equation}\label{eq:dgradeff}
\dgrad^\prime = \dgrad + \clen\tdc(\dlen\tdd)^{-1}\cgrad + \slen\tds(\dlen\tdd)^{-1}\sgrad,
\end{equation}
where $\tdc$, $\tdd$, and $\tds$ are functions of the pulse timings defined in the Appendix, Eq.~\ref{eq:ptimings}. Section~\ref{sec:appcomp} in the Appendix derives Eq.~\ref{eq:dgradeff} and discusses other possible choices for $\dgrad^\prime$.
\item Approximation 3 (A3) uses the Gaussian phase distribution (GPD) approximation to derive models that account explicitly for the butterfly gradients.

\end{itemize}

\subsubsection{Diffusion tensor imaging}

A1 uses the simplest model for DTI, where the signal
\begin{equation}\label{eq:simpledti}
\signal = \signal_0\exp\left(-\dwfactor \ddir^T D \ddir\right),
\end{equation}
\begin{equation}\label{eq:bvalue}
\dwfactor = (\Delta-\dlen/3)(\gyromag\dlen|\dgrad|)^2,
\end{equation}
\begin{equation}\label{eq:bigdelta}
\Delta = \tmix + \dlen + 2\slen + 2\clen + \gapone + \gaptwo,
\end{equation}
$D$ is the DT, $\gyromag$ is the gyromagnetic ratio, $\ddir$ is a unit vector in the direction of $\dgrad$, and $\signal_0$ is the signal with $\dwfactor=0$.

A2 also uses Eq.~\ref{eq:simpledti}, but with $\dgrad^\prime$ from Eq.~\ref{eq:dgradeff} replacing $\dgrad$.

A3 uses the full $\dwfactor$-matrix, analogous to~\cite{MattielloMRM97} for PGSE, rather than the single $\dwfactor$-value in A1 and A2. The Appendix, section~\ref{sec:appdti}, gives the formula.

By assuming a single $\dwfactor$-value, A1 and A2 ignore the cross terms in the $\dwfactor$-matrix, which express the interaction between gradients with different orientation~\cite{MattielloMRM97}. A1 is exact only when $\cgrad=\sgrad=0$. A2 is exact only when $\dgrad$, $\cgrad$ and $\sgrad$ all have the same orientation. A3 accounts for all cross terms so is always exact for Gaussian dispersion assumed in DTI.

\subsubsection{Restricted diffusion}

The GPD approximation to the signal from particles exhibiting restricted diffusion is~\cite{StepisnikPhysicaB93}
\begin{equation}\label{eq:resatt}
\signal = \signal_0 \exp\left( -\frac{\gyromag^2}{\diff^2} \sum_{k=0}^\infty \frac{B_k I_k}{\lambda_k^2}\right),
\end{equation}
where $\diff$ is the free diffusivity within the restricting domain, $B_k$ and $\lambda_k$ are constants that depend on only the geometry of the domain, and $I_k$ depends also on the pulse sequence. For domains with simple geometric shapes such as spheres, separated planes, and cylinders, $B_k$ and $\lambda_k$ have simple analytic form~\cite{StepisnikPhysicaB93}. For PGSE,
\begin{eqnarray}\label{eq:pgsegpd}
I_k & = G_d^2 & (2\dlen\lambda_k^2\diff - 2 + 2Y_k(-\dlen) + 2Y_k(-\Delta) \\\nonumber
&&- Y_k(\dlen - \Delta) - Y_k(-\dlen - \Delta)),
\end{eqnarray}
where $G_d$ is the component of $\dgrad$ in the restricted direction, and $Y_k(x) = \exp(\lambda_k^2 \diff x)$.

Eq.~\ref{eq:pgsegpd} assumes perfectly rectangular diffusion pulses and ignores any diffusion weighting from other pulses. Thus A1 uses Eq.~\ref{eq:pgsegpd} adapted for STEAM by setting $\Delta$ as in Eq.~\ref{eq:bigdelta}.

A2 uses the same formula as A1 with $\dgrad^\prime$ from Eq.~\ref{eq:dgradeff} replacing $\dgrad$.

A3 redefines $I_k$ to accommodate the additional pulses.  Section~\ref{sec:appres} in the Appendix provides the formula.

For restricted diffusion, A1, A2 and A3 are all approximations, since they rely on the GPD approximation. However, A3 accounts for cross terms between the separate pulses, which A1 and A2 ignore.

\subsection{Compensation}\label{sec:comp}

To achieve a particular experiment design, we can compensate for the diffusion weighting of the butterfly gradients using the inverse of approximation A2: for intended gradient vector $\grad$, we acquire instead $\dgrad$ that produces $\dgrad^\prime$ close to $\grad$. For example, directly from Eq.~\ref{eq:dgradeff}, set
\begin{equation}\label{eq:compensation}
\dgrad=\grad - \clen\tdc(\dlen\tdd)^{-1}\cgrad - \slen\tds(\dlen\tdd)^{-1}\sgrad.
\end{equation}
The weightings $\clen\tdc(\dlen\tdd)^{-1}$ and $\slen\tds(\dlen\tdd)^{-1}$ depend only on the timings of the pulses so are constant within one HARDI shell, but may vary between shells or measurements with different $\dwfactor$-value or diffusion time.

\section{Results}\label{sec:experiments}

The central hypothesis is that the new models, A2 or A3, and/or compensation are necessary, because the standard treatment of STEAM diffusion MRI, A1 without compensation, lacks sufficient accuracy.  This section compares signal models A1, A2 and A3, and evaluates the impact of compensation within the context of adapting ActiveAx~\cite{AlexanderNI10} for STEAM. However, we reserve a detailed comparison of STEAM versus PGSE for DTI and/or ActiveAx for future work.

\subsection{ActiveAx protocols}\label{sec:protocols}

The experiments use three imaging protocols.  ActiveAxPGSE is the PGSE ActiveAx imaging protocol from~\cite{DyrbyMRM12} with maximum gradient strength $\gmax=300\,\milli\tesla\meter^{-1}$. ActiveAxSTEAM is a STEAM protocol, also with $\gmax=300\,\milli\tesla\meter^{-1}$, optimised for ActiveAx by adapting the experiment design optimization in~\cite{DyrbyMRM12,AlexanderMRM08} for STEAM. The adaptation simply replaces the estimate of the signal to noise ratio, which is proportional to $\exp(-\te/\ttwo)$ for PGSE and $\exp(-\te/\ttwo)\exp(-\tmix/\tone)$ for STEAM. Table~\ref{table:ActiveAxG300} shows the settings for each of the three HARDI shells that constitute each protocol.  The third protocol, ActiveAxSTEAMCOMP, adapts each $\dgrad$ in ActiveAxSTEAM according to the compensation in section~\ref{sec:comp}.

Every image in ActiveAxSTEAM and ActiveAxSTEAMCOMP has $\clen=1.5\milli\second$, $\cgrad=(0, 0, 0.15)\,\tesla\,\meter^{-1}$, $\slen=1.0\milli\second$, $\sgrad=(0, 0, 0.14)\,\tesla\,\meter^{-1}$, and $\gaptwo=0$.  As an indication of the butterfly gradients' impact, the $\dwfactor$-value from the crushers alone is $250\,\second\,\milli\meter^{-2}$ for the $\dwfactor=3425\,\second\,\milli\meter^2$ shell of ActiveAxSTEAM, in contrast to $10\,\second\,\milli\meter^{-2}$ for ActiveAxPGSE. Since $\cgrad$ and $\sgrad$ are both along the slice direction $(0, 0, 1)$, the compensation $\grad - \dgrad$, from Eq.~\ref{eq:compensation}, is along the negative slice direction; $\|\grad - \dgrad\|_2=43.4\,\milli\tesla\,\meter^{-1}$, $68.5\,\milli\tesla\,\meter^{-1}$ and $76.0\,\milli\tesla\,\meter^{-1}$, for the three shells, respectively. To illustrate practical implementation of ActiveAxSTEAMCOMP, the first gradient direction in the $\dwfactor=3425\,\second\,\milli\meter^2$ shell is $[0.85, 0.48, 0.23]$ and $\|\dgrad\|_2=113.5\,\milli\tesla\,\meter^{-1}$, so the intended gradient vector is $[95.9, 54.4, 26.6]\,\milli\tesla\,\meter^{-1}$. Eq.~\ref{eq:compensation}, tells us to type $\dgrad=[95.9, 54.4, -41.9]\,\milli\tesla\,\meter^{-1}$ ($\ddir=[0.81, 0.46, -0.35]$ and $\|\dgrad\|_2=118\,\milli\tesla\,\meter^{-1}$) into the scanner console instead.

\begin{figure}
\includegraphics[width=\linewidth]{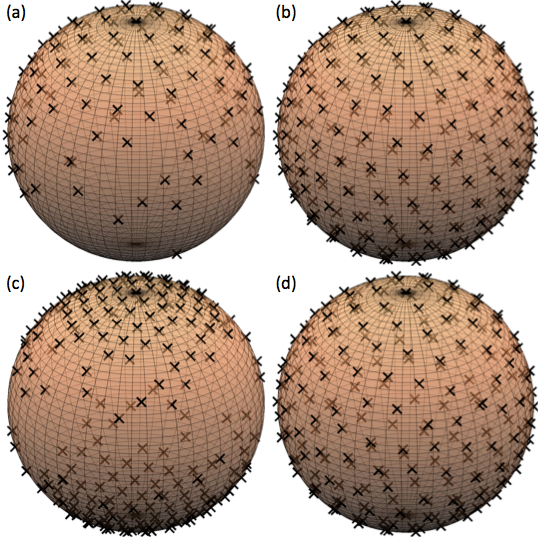}
\caption{\label{fig:GradientDirections}Illustration of the target and effective gradient directions in the STEAM protocols using the 108 directions in the $\dwfactor=3425\,\second\,\milli\meter^{-2}$ shell of the ActiveAxSTEAM protocol. A black cross marks each direction; shaded crosses are on the far side of the sphere. Panel (a) is the target set of gradient directions. Panel (b) shows the target set with a cross in both the positive and negative gradient direction to show the isotropic distribution more clearly.  Panel (c) shows the set of effective gradient directions, i.e. the direction of $\dgrad^\prime$ in Eq.~\ref{eq:dgradeff}, without compensation (ActiveAxSTEAM); they skew strongly towards the slice direction. Panel (d) shows the effective gradient directions after compensation (ActiveAxSTEAMCOMP), which are close to the target set.}
\end{figure}

Figure~\ref{fig:GradientDirections} shows the distribution of effective gradient directions, i.e. the orientation of $\dgrad^\prime$ from Eq.~\ref{eq:dgradeff}, for the $\dwfactor=3425\second\,\milli\meter^{-2}$ shell of ActiveAxSTEAM and ActiveAxSTEAMCOMP to illustrate the disruption to the HARDI design. Without compensation, the butterfly gradients skew the effective gradient directions strongly towards the slice direction.  The compensated protocol has evenly distributed effective gradient directions.

\subsection{Data acquisition}\label{sec:dataacqn}

We acquire data from a fixed monkey brain, prepared as in~\cite{DyrbyHBM11}, using all three protocols in a single contiguous session. The live monkey was handled and cared for on the Island of St. Kitts according to a protocol approved by the local ethics committee (The Caribbean Primate Center of St. Kitts). The image volume is $256\times128$ voxels in plane with 15 contiguous sagittal slices including the mid-sagittal plane; voxels are $0.5\milli\meter$ isotropic. 

The ActiveAxSTEAMCOMP acquisition has two imperfections. First, the butterfly gradients affect the nominal $\dwfactor=0$ images, as well as the diffusion weighted images. In theory, the compensation works for them too by adding non-zero $\dgrad$ in the negative slice direction.  However, imperfect r.f. pulses prevent use of the compensation for the nominal $\dwfactor=0$ images in practice. In the absence of a strong diffusion gradient, the compensation counteracts the effect of the crusher gradients, allowing additional echoes to affect the signal and leading to severe image artifacts. Thus the nominal $\dwfactor=0$ images remain uncompensated with $\dgrad=\vect{0}$.

\begin{figure}
\includegraphics[width=\linewidth]{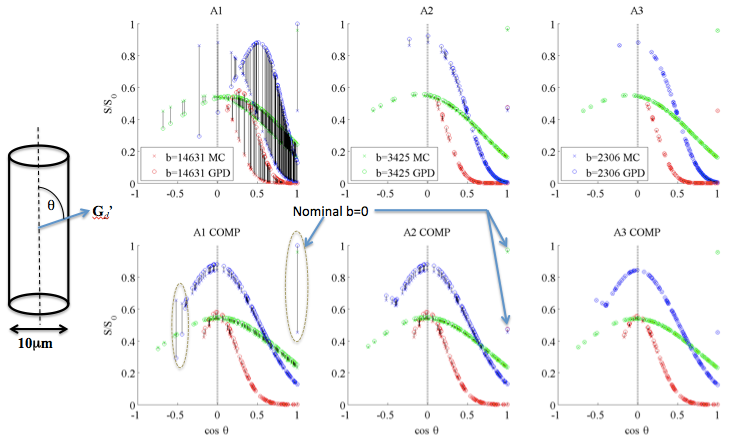}
\caption{\label{fig:RestrictedSims}Comparison of the STEAM signal estimates using A1, A2 and A3 from particles restricted within an impermeable cylinder of diameter $10\micro\meter$ with ground-truth signals from MC simulation. Each panel plots normalized signals, $\signal/\signal_0$, against $\cos\theta$ where $\theta$ is the angle between the effective gradient direction $\dgrad^\prime$ (Eq.~\ref{eq:dgradeff}) and the cylinder axis (positive slice direction). A black line connects each estimate with the corresponding ground truth. The top row shows plots for the ActiveAxSTEAM protocol. The bottom row shows equivalent plots for ActiveAxSTEAMCOMP. The dotted vertical line indicates the perpendicular gradient orientation where we expect the largest signal. The ellipses in the bottom left figure highlight the imperfectly compensated measurements.}
\end{figure}

The second imperfection occurs in a small number of measurements for which the slice-direction components of $\dgrad$ after compensation exceed $\gmax$.  The scanner automatically truncates that component at $\gmax$, so the effective gradient vector departs from what the compensated protocol intends.  The second imperfection is avoidable by negating the original gradient direction before compensation. However, we retain the imperfection here, as it (a) affects only two measurements significantly (both in the $\dwfactor=2306\,\second\,\milli\meter^{-2}$ shell; see figure~\ref{fig:RestrictedSims} later) and (b) helps to illustrate differences between A1 and A2 (figure~\ref{fig:RestrictedSims}). 

\subsection{DTI}\label{sec:dtiexperiments}

This section evaluates bias in the DT estimated using A1, A2 and A3 from both compensated and uncompensated acquisition.  Simulation experiments quantify the effects in idealised conditions.  Experiments with the monkey brain data confirm the trends on measured data.  Both experiments focus on the $b=3425\second\,\milli\meter^{-2}$ shell from the ActiveAxSTEAM protocol, which has $\dwfactor$-value typical for ex-vivo DTI~\cite{DyrbyHBM11} and long $\tmix$ that exploits the benefits of STEAM, but also emphasises the diffusion weighting of the butterfly gradients. 

\subsubsection{Simulations}

{\bf Experiment.} The synthetic data do not reflect the two imperfections in the scanner data, so the nominal $\dwfactor=0$ measurements are compensated, and no truncation of the gradient vectors at $\gmax$ occurs (no measurements in the $b=3425\second\,\milli\meter^{-2}$ shell are affected by this anyway).  Eqs.~\ref{eq:gaussian} and~\ref{eq:bmatrix} in the Appendix provides synthetic data from the DT model.  

The experiments use two DTs, one with eigenvalues $\{0.6, 0.2, 0.2\}\times10^{-9}\,\meter^2\second^{-1}$, which are typical of coherent white matter in fixed brain tissue at this $\dwfactor$ value, and the other $\{0.4, 0.4, 0.4\}\times10^{-9}\,\meter^2\second^{-1}$, which is isotropic with trace typical of grey matter~\cite{DyrbyHBM11}.  The anisotropic DT has two variations: the first has principal eigenvector $\pd=[0, 0, 1]$, so that $\cgrad$ and $\sgrad$ are parallel to the fibre direction, and the second has $\pd=[1, 0, 0]$, so they are perpendicular.

Each experiment adds Rician noise so that the signal to noise ratio of the unweighted signal is $20$.  Weighted linear least squares fitting~\cite{JonesMRM04} estimates the DT using each approximation from which we compute the eigenvalues, fractional anisotropy (FA) and $\pd$.  We repeat the procedure over $10000$ independent noise trials and compute the mean and standard deviation of the largest eigenvalue $\lambda_1$ and the FA.  We also compute the mean angle $\alpha$ between the estimated and true $\pd$, for the anisotropic DTs. For all DTs, we compute the direction concentration $\dirconc = -\log(1-E)$, where $E$ is the largest eigenvalue of the mean dyadic tensor~\cite{AlexanderNI05}. The direction concentration is zero for an isotropic set of directions and increases as the variance of the distribution decreases, reaching infinity when all align perfectly. Typical values of $\dirconc$ for similar noise trials with anisotropic tensors in~\cite{AlexanderNI05} are 6 to 8.  Unbiased noise trials with the isotropic tensor should produce $\dirconc$ close to zero.

To give some idea of the significance of the effects in a human imaging protocol, we repeat the experiment using in-vivo settings for a 3T clinical scanner.  The protocol has $7$ nominal $\dwfactor=0$ images and $60$ gradient directions with $\dwfactor=1007\,\second\,\milli\meter^{-2}$; $\|\dgrad\|_2=40\,\milli\tesla\,\meter^{-1}$, $\tmix=120\,\milli\second$, $\dlen=8\,\milli\second$, $\gapone=\gaptwo=0$, $\clen=0.5\,\milli\second$, $\slen=5.5\,\milli\second$, $\cgrad=[20,20,20]\,\milli\tesla\,\meter^{-1}$ and $\sgrad=[0,0,6]\,\milli\tesla\,\meter^{-1}$.  The butterfly gradients are weaker than the ex-vivo protocol, because the voxel size is larger ($2.3\,\milli\meter$ isotropic). The test DTs have eigenvalues $\{1.7, 0.2, 0.2\}\times10^{-9}\,\meter^2\,\second^{-1}$ and $\{0.7, 0.7, 0.7\}\times10^{-9}\,\meter^2\,\second^{-1}$.

{\bf Results.} Tables~\ref{table:AnisoG300} and~\ref{table:IsoG300} list statistics for the fixed-tissue simulations with the anisotropic and isotropic DTs, respectively. Note that perfect compensation makes A1 and A2 equivalent.

Without compensation, A1 shows significant bias in FA, $\lambda_1$ and $\pd$ with both orientations of the anisotropic DT.  Bias is most severe for $\pd$ parallel to the butterfly gradients where the DT estimation completely fails. Estimates of the isotropic DT show artifactual non-zero FA and significant direction concentration: $\dirconc=1.5$ means $95\%$ of directions are within $6^\circ$ of the mean. Compensation dramatically improves A1. Some downward bias remains in both FA and $\lambda_1$ of the anisotropic DTs, but the bias is similar for both orientations. Compensation largely removes artifactual non-zero FA and orientational bias in the isotropic DT estimates: $\dirconc=0.4$ is typical for a uniformly distributed random sample of $10000$ directions and the $95\%$-angle is over $25^\circ$. 

Without compensation, A2 and A3 produce very similar results. Both significantly reduce bias compared to A1, although bias remains orientationally dependent and is strongest with parallel $\cgrad$ and $\sgrad$.  Compensation reduces bias and variance of parameter estimates from A3, especially for parallel $\pd$, and removes orientational dependence. With compensation, A3 shows no benefit over A1 or A2. 

Tables~\ref{table:AnisoHuman} and~\ref{table:IsoHuman} show corresponding results from the in-vivo human protocol. Without compensation, A1 still produces considerable bias, which A2 or A3 reduces. The compensation provides only minor further improvements with A3.  

{\bf Conclusions.} Two separate effects cause unnecessary bias in the parameter estimates: model inaccuracy and disrupted experiment design. Model inaccuracy is the dominant cause of the large bias from A1 without compensation. The large reduction in bias from replacing A1 with A2 or A3 demonstrates the importance of accounting for the butterfly gradients in the model. The lack of performance difference between A2 and A3 shows that the cross terms in the $\dwfactor$-matrix are negligible.

The bias we observe in FA and $\lambda_1$ from A3 with compensation is unavoidable, since the model is exact and the experiment design is not disrupted.  It comes from Jones' ``squashed-peanut'' effect~\cite{JonesMRM04}: a Rician noise effect as measurements with gradient parallel to $\pd$ approach the noise floor. Differences in results from A3 with and without compensation show the effect of the experiment design disruption. The disruption to the experiment design affects parameter estimates most strongly with parallel butterfly gradients, because the additional diffusion weighting in the fibre direction pushes parallel signals further into the noise floor. Compensation reduces bias and improves precision by removing the experiment design disruption, which also removes the orientational dependence of the bias and variance. 

The performance differences are less marked in the human protocol, because the butterfly gradients are smaller. However, values of $\alpha$ between $2.5^\circ$ and $12^\circ$ that we observe for A1 without compensation are at least as large as orientational bias incurred by failing to account for small head motions in the $\dwfactor$-matrix, which~\cite{LeemansMRM09} finds sufficient to disrupt tractography.

\subsubsection{Monkey data}

{\bf Experiment.} We fit the DT to the $b=3425\second\,\milli\meter^{-2}$ shell of ActiveAxSTEAM and ActiveAxSTEAMCOMP, as well as the $b=3084\second\,\milli\meter^{-2}$ shell of ActiveAxPGSE, using weighted linear least squares and construct colour-coded $\pd$ maps~\cite{PajevicMRM99}.  We quantify the orientational similarity between pairs of DT volumes by computing the mean over the brain of the absolute dot product of principal directions weighted by DT linearity~\cite{WestinMICCAI99}. 

\begin{figure}
\includegraphics[width=\linewidth]{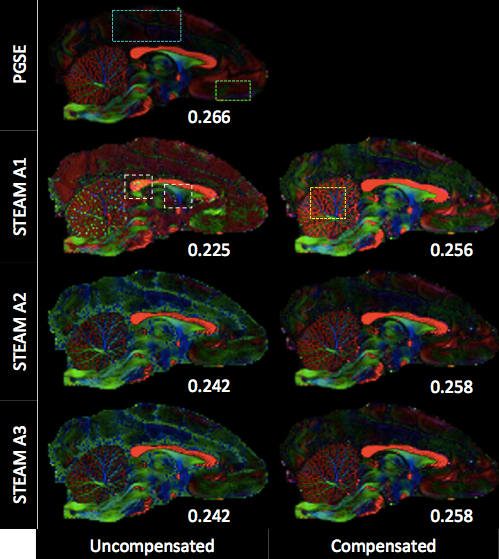}
\caption{\label{fig:BrainDTI}Direction-encoded colour maps~\cite{PajevicMRM99} for the mid-sagittal slice of the monkey brain from the $\dwfactor=3084\,\second\,\milli\meter^{-2}$ shell of ActiveAxPGSE (top left), the $\dwfactor=3425\,\second\,\milli\meter^{-2}$ shell of ActiveAxSTEAM (left) and ActiveAxSTEAMCOMP (right). Rows 2-4 show the maps reconstructed with A1, A2 and A3, respectively. The numbers quantify the orientational similarity (definition in the text) between each STEAM map and the PGSE map.  The number in the PGSE panel is the similarity of the maps from the $\dwfactor=3084\,\second\,\milli\meter^{-2}$ and $\dwfactor=2243\,\second\,\milli\meter^{-2}$ shells of ActiveAxPGSE.}
\end{figure}

{\bf Results.} Figure~\ref{fig:BrainDTI} compares maps from PGSE with STEAM for each approximation qualitatively. The number next to each STEAM map is the orientational similarity with PGSE; higher numbers show greater agreement. The number next to the PGSE map is the orientational similarity of the $b=2243\second\,\milli\meter^{-2}$ and $b=3084second\,\milli\meter^{-2}$ shells of ActiveAxPGSE. 

For ActiveAxSTEAM, A1 introduces upward bias in FA in the superior half of the brain where diffusion should be close to isotropic, such as the area in the cyan box on the PGSE map.  The maps also show orientation bias towards the slice direction, which is left-right in the brain (the map appears red). The white boxes show bias in anisotropic regions: the left box shows severely biased orientation estimates (some voxels appear green rather than red) in the corpus callosum, where the fibres are parallel to the butterfly gradients; the right box shows less biased orientation estimates in the fornix, which has perpendicular fibres. A2 and A3 are qualitatively indistinguishable from one another and are more consistent with the PGSE map than A1, e.g. in the white boxes. However, they still show upward bias in FA together with consistent artifactual orientation in isotropic regions (blue/green colour in cyan box region). 

For ActiveAxSTEAMCOMP, all maps appear more similar to PGSE than the uncompensated maps and have low FA in the cyan box region. A2 and A3 are indistinguishable. The compensated A1 map shows generally higher anisotropy, for example in the cerebellum marked by the yellow box. Some differences in orientation between the compensated STEAM and PGSE maps still appear, for example in the area marked by the green box.

{\bf Conclusions.} Differences among the maps broadly reflect the bias we observe in the simulation experiment. A1 uncompensated shows artifactual raised FA in isotropic regions and orientation bias towards left-right arises from the butterfly gradients enhancing attenuation in that direction. Moreover, the white boxes demonstrate the orientational dependence of the bias: as in the simulations it is most significant when $\pd$ and butterfly gradients are parallel.

Compensation generally reduces bias. Differences between A1 and A2 with compensation appear because A1 does not account for the imperfections in the compensation. Maps from ActiveAxSTEAMCOMP with A3 and PGSE do not match perfectly, because the diffusion times and $\dwfactor$-values differ.

The trends in quantitative orientational similarity confirm the intuition from the qualitative maps.

\subsection{Restricted diffusion}\label{sec:resexperiments}

This section uses the full ActiveAx data sets to demonstrate the models and compensation in an application that exploits restricted diffusion. 

\subsubsection{Simulations}

The simulation experiment compares the accuracy of A1, A2 and A3 for restricted diffusion in a cylinder. 

{\bf Experiment.} The Monte-Carlo (MC) diffusion simulation system from~\cite{HallTMI09}, implemented in the Camino toolkit~\cite{Camino}, provides synthetic ground truth measurements accounting precisely for all gradient pulses and timings. The simulations use ActiveAxSTEAM and ActiveAxSTEAMCOMP (from table~\ref{table:ActiveAxG300}) and this time include the imperfections in the ActiveAxSTEAMCOMP scanner data. Each simulation uses $160000$ walkers and $5000$ timesteps, which produces unbiased synthetic measurements with standard deviation less than $10^{-4}\signal_0$ \cite{HallTMI09}. All the walkers are trapped inside an impermeable cylinder (no extra-axonal contribution) with diameter $10\,\micro\meter$ and axis aligned with the slice direction; free diffusivity is $600\,\micro\meter^2\second^{-1}$. 

{\bf Results.} Figure~\ref{fig:RestrictedSims} compares the synthetic data from the MC simulation with predictions from A1, A2 and A3 for ActiveAxSTEAM (top row) and ActiveAxSTEAMCOMP (bottom row). 

Without compensation, A1 shows large departures from the ground truth MC signals. A2 and A3 match the MC data much more closely: the maximum error against corresponding MC data points is 0.6 for A1, 0.04 for A2, and 0.01 for A3.  

With compensation, large departures in A1 remain only in measurements with truncated gradient vectors (the two $\dwfactor=2306\,\second\,\milli\meter^{-2}$ measurements with the most negative z-component in the left ellipse) and the uncompensated nominal $\dwfactor=0$ measurements (right ellipse). A2 is equivalent to A1 apart from the imperfectly compensated measurements, which A2 predicts closely. A3 matches the MC data slightly better than A2, in particular for the $\dwfactor=2306\,\second\,\milli\meter^{-2}$ shell (blue). A3 shows small departures from the MC data, especially in the high $\dwfactor$-value shell (red).

{\bf Conclusions.} A1 uncompensated predicts the signal poorly.  In particular, ignoring the butterfly gradients predicts the highest signal to occur when $\dgrad$ is along the cylinder axis, whereas the peak actually occurs when $\dgrad^\prime$ is perpendicular ($\cos\theta=0$ in the figure). 

In contrast to the free diffusion experiments, A3 does not provide exact predictions.  Departures from the ground truth arise from violation of the GPD assumption. The departures reduce as cylinder diameter  decreases. However, A3 does provide a benefit over A2 showing that cross terms are influential for restricted diffusion.  The benefit also reduces as diameter decreases.

\subsubsection{Monkey data}

{\bf Experiment.} We fit the minimal model of white matter diffusion (MMWMD)~\cite{AlexanderNI10,DyrbyMRM12} to the full data acquisition from each protocol in table~\ref{table:ActiveAxG300} using the procedure outlined in~\cite{AlexanderNI10}.  The mixing time varies among the different shells in the STEAM protocols, so we first estimate $\tone$ from the nominal $\dwfactor=0$ images and fix its value for the subsequent MMWMD fitting.

\begin{figure}
\includegraphics[width=\linewidth]{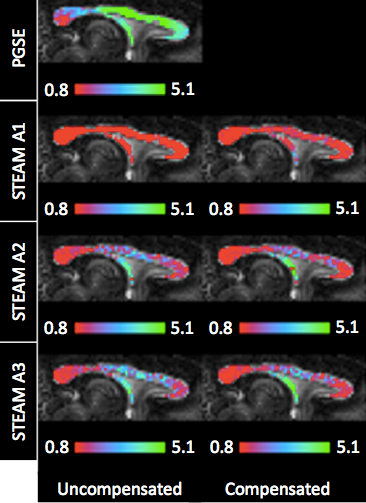}
\caption{\label{fig:AxonDiameterIndex}Maps of the axon diameter index~\cite{AlexanderNI10} over the mid-sagittal corpus callosum recovered from the ActiveAxPGSE data (top left), ActiveAxSTEAM (left) and ActiveAxSTEAMCOMP (right) with each approximation. }
\end{figure}

{\bf Results.} Figure~\ref{fig:AxonDiameterIndex} shows the axon diameter index maps from ActiveAxPGSE, and ActiveAxSTEAM and ActiveAxSTEAMCOMP with each approximation. The axon diameter index map from ActiveAxPGSE shows the familiar high-low-high trend from splenium through mid-body to genu, as in previous applications of ActiveAx~\cite{AlexanderNI10,DyrbyMRM12}. 

The maps from A1 show no clear trend for either ActiveAxSTEAM or ActiveAxSTEAMCOMP. However, all maps from A2 and A3 show the high-low-high trend, although the axon diameter index itself is consistently lower from STEAM than PGSE. A3 provides a greater range of axon diameter index and reproduces the trend more clearly than A2.  Fitting errors (not shown) are significantly larger for A1 than either A2 or A3 and slightly lower for A3 than A2. 

{\bf Conclusions.} Severe model inaccuracy in A1 prevents sensible estimates of the axon diameter index;  with ActiveAxSTEAMCOMP the imperfectly compensated measurements disrupt MMWMD fitting. A2 captures the imperfectly compensated measurements allowing the usual trend to emerge. Visible differences from A2 to A3 reflect lesser accuracy in A2, which the simulations demonstrate. Although we cannot verify directly that compensation and A3 produce better results, differences appearing among maps in figure~\ref{fig:AxonDiameterIndex} suggests that both are necessary.

Lower axon diameter index from STEAM compared to PGSE is somewhat counterintuitive, because longer diffusion times in STEAM increase sensitivity to larger axons over PGSE~\cite{DyrbyMRM12}. Thus we might expect the axon diameter index to increase.  However, if no large axons are present, the STEAM data provide better information to reject any likelihood of their existence. This reduces the tails of the posterior distribution in the large diameter range that the PGSE data may permit, reducing the axon diameter index, which is the mean of the posterior~\cite{AlexanderNI10}. Indeed, the axon diameter indices from A3 compensated are closer to the values we might expect~\cite{AlexanderNI10} based on histology~\cite{LamantiaJCompNeuro90} than those from PGSE. However, that histology is from the brain of a different species, so further work is required to confirm this hypothesis.

\section{Discussion}

This paper highlights the need to account for the diffusion weighting of butterfly gradients in STEAM diffusion MRI.  We provide signal models for both free and restricted diffusion that accommodate their effect. We also introduce a simple compensation to the acquired diffusion gradient that minimizes disruption to the experiment design the butterfly gradients cause.  DTI and ActiveAx experiments with both synthetic and fixed monkey-brain data illustrate the potential for severe bias from ignoring the butterfly gradients (A1 uncompensated, the usual approach) and the major benefits of our improved models in avoiding unnecessary bias. They show further that retaining experiment design, in particular in HARDI applications, through our compensation further improves accuracy and precision of parameter estimates and avoids orientational dependence of both.

\subsection{Recommendations}

For acquisition, we recommend the compensation wherever possible in STEAM diffusion MRI, as it has no cost in terms of acquisition or post-processing. The compensation is straightforward to implement: it requires no pulse programming, simply adjustments to the scheme file specifying the gradient strengths and directions to the scanner. However, users should check how the scanner truncates or normalises gradient vectors to avoid the imperfections we mention in section~\ref{sec:dataacqn}. The compensation is particularly important in HARDI methods, but single-direction model-based STEAM diffusion MRI applications, such as~\cite{AssafMRM08,OngNIMG10}, are also likely to benefit significantly.

For data analysis, we strongly recommend avoiding A1, whereas A2 is sufficient for many practical circumstances, such as DTI. Although, A1 and A2 are equivalent in theory if the acquisition uses compensation, imperfect r.f. pulses sometimes prevent the compensation of low $\dwfactor$-value measurements in practice, making A2 necessary. For models involving restricted diffusion, the slightly more accurate A3, which has a cost of about double the computation time, appears beneficial over A2.

\subsection{Limitations and alternatives}

The GPD approximation for restricted diffusion generally provides a reasonable approximation for the range of $\dwfactor$-values and cylinder diameters relevant to the applications of interest here~\cite{BalinovJMRA93,IanusISMRM12}.  However, it breaks down in some signal regimes; for example, it does not capture the characteristic $q$-space diffraction patterns in the restricted diffusion signal~\cite{CallaghanNature91}. These circumstances require more precise estimates of the signal for example from Callaghan's matrix formulation~\cite{CallaghanJMR97} and related numerical techniques that extend the idea to three-dimensions~\cite{GrebenkovRevModPhys07,DrobnjakJMR11}.

The idealized pulse sequence model we use assumes zero ramp time for all pulses. The GPD method extends easily to accommodate non-zero ramp times~\cite{IanusISMRM12}, although in most practical situations they have little effect on the signal estimate.

The butterfly gradients in our preclinical ex-vivo application are stronger than in most in-vivo human applications, because the image slices are thinner. The strong gradients emphasize the disruption of the intended experiment design; the effect is less marked in in-vivo human applications, as tables~\ref{table:AnisoHuman} and~\ref{table:IsoHuman} show. However, even small biases can disrupt subsequent analysis, such as tractography~\cite{LeemansMRM09}, so the methods we propose are still necessary.

We do not consider additional diffusion weighting from other imaging gradients, such as echo-planar imaging (EPI) gradient trains, which~\cite{MattielloMRM97} demonstrate can be significant. Our data acquisition does not use EPI, so such contributions are irrelevant here. However, the general modelling and compensation approach extends naturally to account for these gradients if necessary. We also do not consider background gradients, which~\cite{CottsJMR89,FinsterbuschJMR08} design versions of the STEAM pulse sequence to compensate for. Our compensation and models adapt naturally for those sequences and future work will study the necessity for such adaptations in brain-imaging applications.

The one-sidedness of the set of gradient directions affects the amount of bias that the butterfly gradients introduce to fitted parameters. Figure~\ref{fig:GradientDirections}(a) shows that most of the directions in our protocols have positive z-component, so the butterfly gradients skew them away from the slice plane. Conversely, they skew most directions with negative z-component towards the slice plane.  An even distribution of \emph{signed} directions could reduce bias in parameter estimates, because the errors for positive and negative directions cancel to some extent. However, we do not recommend this solution, as it produces large fitting errors and is likely to mask undesirable effects on estimated parameters.

Other strategies for avoiding the effects of the butterfly gradients include simply turning the crushers off in the diffusion weighted measurements and relying on the diffusion gradients to crush unwanted echoes~\cite{DhitalISMRM10}. This requires sufficiently high diffusion weighting and is generally not possible for the nominal $\dwfactor=0$ images that most protocols require for normalization; the models we propose are essential for explaining the signal in those images. Moreover, the slice-select gradients are always necessary in imaging applications.

\subsection{Conclusions}

We demonstrate here that imaging gradients in the STEAM sequence can severely disrupt HARDI experiment design and cause bias in parameter estimates if ignored.  The models and methods we present solve these problems and enable widespread uptake of STEAM diffusion MRI. They allow future work to evaluate and exploit the potential benefits of STEAM especially for diffusion MRI on high field scanners where low $\ttwo$ prevents long diffusion time PGSE. In particular, they enable us to evaluate STEAM ActiveAx for better sensitivity to large axons, which is the focus of our current work.

\section*{Acknowledgements}

We thank Prof Maurice Ptito (University of Montreal and Copenhagen University) and Mark Burke (Howard University) for providing the fixed monkey brain. Both authors acknowledge the support of Eurpoean Commission Framework 7 through the CONNECT consortium. The EPSRC support DCA's work on this topic with grants EP/H046410/01 and EP/E007748. TBD was also supported by the Lundbeck foundation.

\appendix

\section{Appendix}

\subsection{Gaussian dispersion and DTI}\label{sec:appdti}

On the assumption of zero-mean Gaussian particle dispersion (the DT model), the general formula~\cite{PriceConcMR97}
\begin{equation}\label{eq:gaussian}
\signal = \signal_0 \exp (-B\cdot D),
\end{equation}
predicts the signal, where $B = \gyromag^2 \int \vect{F}(t) \vect{F}^T(t) \rmd t$ is the $\dwfactor$-matrix~\cite{MattielloMRM97},
\begin{equation}\label{eq:ft}
\vect{F}(t) = \int_0^t \grad(t)dt,
\end{equation}
$\grad(t)$ is the gradient vector at time $t$, and $\cdot$ is the matrix scalar product.

For idealised PGSE or STEAM with $\cgrad=\sgrad=0$, i.e. approximations A1 or A2, Eq.~\ref{eq:gaussian} reduces to Eq.~\ref{eq:simpledti}. However, for the full pulse sequence outlined in section~\ref{sec:sequence}, i.e. approximation A3,
\begin{eqnarray}\label{eq:bmatrix}
B & = & \dlen^2\tdd\dgrad\dgrad^T + \clen^2\tcc\cgrad\cgrad^T + \slen^2\tss\sgrad\sgrad^T\\\nonumber
&& + \dlen\clen\tdc(\dgrad\cgrad^T + \cgrad\dgrad^T) + \dlen\slen\tds(\dgrad\sgrad^T+\sgrad\dgrad^T)\\\nonumber
&& + \clen\slen\tcs(\cgrad\sgrad^T+\sgrad\cgrad^T),
\end{eqnarray}
where
\begin{eqnarray}\label{eq:ptimings}
\tdd & = & \gapone + \gaptwo + \tmix + 2\clen + 2\dlen/3 + 2\slen,\\\nonumber
\tcc & = & \tmix + 2\clen/3 + 2\slen,\\\nonumber
\tss & = & \tmix + 2\slen/3,\\\nonumber
\tdc & = & \tmix + \clen + 2\slen,\\\nonumber
\tds & = & \tmix + \slen,\\\nonumber
\tcs & = & \tmix + \slen.
\end{eqnarray}

The expression in Eq.~\ref{eq:bmatrix} is the sum of pairwise interactions between the diffusion, crusher and slice-select pulses, similar to the $\dwfactor$-matrix for PGSE in~\cite{MattielloMRM97}.

In the absence of any diffusion weighting, the STEAM signal in terms of a steady-state magnetization $\ssmag$ and relaxation constants $\tone$ and $\ttwo$ is
\begin{equation}\label{eq:relaxation}
\signal_0(\te, \tr, \tmix) = \ssmag (1 - \exp(-(\tr-\tmix)/\tone)) \exp(-\tmix/\tone) \exp(-\te/\ttwo),
\end{equation}
where $\te$ is the echo time and $\tr$ is the repetition time.  Thus, in general we require knowledge of, or must estimate, $\tone$ and $\ttwo$ to estimate diffusion parameters. Normally, $\tr\gg\tmix$ and $\tr\gg\tone$, so that $\exp(-(\tr-\tmix)/\tone)\approx0$ and substituting Eq.~\ref{eq:relaxation} into Eq.~\ref{eq:gaussian} and taking logs gives
\begin{eqnarray}\label{eq:logsig}
\log \signal & = & \log \ssmag - \tmix/\tone - \te/\ttwo - B\cdot D.
\end{eqnarray}
Thus, we can obtain linear estimates of $\log \ssmag$, $\tone$, $\ttwo$ and $D$ simultaneously given a set of measurements with sufficiently diverse $B$, $\tmix$ and $\te$.  Specifically, $A = X^\star L$, where
\begin{equation}
A^T = (\log\ssmag, 1/\tone, 1/\ttwo, D_{xx}, D_{xy}, D_{xz}, D_{yy}, D_{yz}, D_{zz})
\end{equation}
contains all the parameters to estimate, $L^T = (\log\signal_1, \log\signal_2, \cdots, \log\signal_N)$ contains all the log signals, and $X^\star$ is the pseudoinverse of the design matrix $X$, which has rows
\begin{equation}
(1, -\tmix, -\te, -B_{xx}, -B_{xy}, -B_{xz}, -B_{yy}, -B_{yz}, -B_{zz}).
\end{equation}

 Single-shell HARDI protocols can keep $\te$, $\tr$, and $\tmix$ constant to avoid having to estimate $\tone$ or $\ttwo$. For multiple $\dwfactor$-values often we can keep $\te$ constant, but $\tmix$ needs to vary to retain the short-$\te$ benefits of STEAM. Thus we can ignore $\ttwo$, but need to estimate $\tone$. For fixed $\tmix$ or $\te$, we remove the second or third, respectively, element of $A$ and column of $X$.

\subsection{Compensation}\label{sec:appcomp}

The simple correction for $\dgrad$ in Eq.~\ref{eq:compensation} to compensate for the butterfly gradients sets $\dgrad-\grad$ to the linear combination $g\cgrad+h\sgrad$ that minimises diffusion weighting in the nominal $\dwfactor=0$ images, i.e. when the intended $\grad=\vect{0}$. Eq.~\ref{eq:compensation} comes from setting $\dgrad=g\cgrad+h\sgrad$ and minimising the trace of the $\dwfactor$-matrix in Eq.~\ref{eq:bmatrix} with respect to $g$ and $h$ to obtain $g=-\clen\tdc(\dlen\tdd)^{-1}$ and $h=-\slen\tds(\dlen\tdd)^{-1}$.  The approximation A2 in Eq.~\ref{eq:dgradeff} simply inverts the compensation to obtain $\dgrad^\prime$.

Another choice of $\dgrad^\prime$ for A2 is $(b_1/(\dlen^2\tdd))^\frac12\vect{v}_1$, where $\vect{v}_1$ is the primary eigenvector of the $\dwfactor$-matrix and $b_1$ is the corresponding eigenvalue. However, the two choices for $\dgrad^\prime$ are very similar in practice and the former is simpler to compute. The maximum difference between the two $\dgrad^\prime$ is around $1\%$ over the whole ActiveAxSTEAM protocol.

In practice, $\sgrad$ must be in the slice direction, but $\cgrad$ can have any orientation. Both choices of $\dgrad^\prime$ accommodate arbitrary and separate orientations of $\cgrad$ and $\sgrad$.

\subsection{Restricted diffusion}\label{sec:appres}

For the full sequence outlined in section~\ref{sec:sequence},
\begin{equation}
I_k = 
s_{ddk}G_d^2 + s_{cck}G_c^2 + s_{ssk}G_s^2 + s_{dck}G_dG_c + s_{dsk}G_dG_s + s_{csk}G_cG_s
\end{equation}
where $G_d$, $G_c$ and $G_s$ are the components of $\dgrad$, $\cgrad$ and $\sgrad$, respectively, in the restricted direction, and
\begin{eqnarray}
s_{ddk} & = & -2 + 2Y_k(-\dlen) - Y_k(-(\gapone + \gaptwo + 2 \clen + 2 \slen + \tmix)) \\\nonumber
&&+ 2Y_k(-(\gapone + \gaptwo + 2 \clen + 2 \slen + \dlen + \tmix)) - \\\nonumber
&&Y_k(-(\gapone + \gaptwo + 2\clen + 2\slen + 2\dlen + \tmix)) + 2 \lambda_k^2 D \dlen,\\
s_{dck} & = & (Y_k(\gapone) + Y_k(\gaptwo)) (Y_k(\clen) - 1) (Y_k(\dlen) - 1) \times\\\nonumber
&&(Y_k(\clen + 2 \slen + \tmix) - 1) Y_k(-(\gapone + \gaptwo + 2 \clen + 2 \slen + \dlen + \tmix)),\\
s_{dsk} & = & (Y_k(\gapone) + Y_k(\gaptwo)) (Y_k(\slen) - 1) (Y_k(\dlen) - 1) (Y_k(\slen + \tmix) - 1) \times\\\nonumber
&&Y_k(-(\gapone + \gaptwo + \clen + 2 \slen + \dlen + \tmix)),\\
s_{cck} & = & -2 + 2Y_k(-\clen) - Y_k(-(2 \slen + \tmix)) + 2Y_k(-(\clen + 2 \slen + \tmix)) \\\nonumber
&&- Y_k(-(2 \clen + 2 \slen) + \tmix)) + 2 \lambda_k^2 D \clen,\\
s_{csk} & = & 2 (Y_k(\clen) - 1) (Y_k(\slen) - 1) (Y_k(\slen + \tmix) - 1) \times\\\nonumber
&&Y_k(-(\clen + 2 \slen + \tmix)),\\
s_{ssk} & = & -2 + 2Y_k(-\slen) - Y_k(-\tmix) + 2Y_k(-(\slen + \tmix)) \\\nonumber
&&- Y_k(-(2 \slen + \tmix)) + 2 \lambda_k^2 D \slen.
\end{eqnarray}
To evaluate these expressions, expand them before implementing to avoid numerical problems; code is available from the authors on request.

\bibliography{DiffusionMRI}

\clearpage

\begin{table}
\begin{tabular}{||l|l|l|l|l|l|l|l||}
\hline
$N$ & $K$ & $|\dgrad|/\milli\tesla\,\meter^{-1}$   & $\Delta/\milli\second$   & $\dlen/\milli\second$   & $\dwfactor/\second\,\milli\meter^{-2}$ & $\te/\milli\second$ & $\tr/\milli\second$ \\
\hline
103 & 25 & 300.0 & 12.9 & 5.6 & 2243 & 36.8 & 2600 \\

106 & 25 & 219.2 & 20.4 & 7.0 & 3084 & 36.8 & 2600 \\

80 & 25 & 300.0 & 18.8 & 10.5 & 10838 & 36.8 & 2600 \\
\hline
\end{tabular}
\begin{center}(a)\end{center}
\begin{tabular}{||l|l|l|l|l|l|l|l|l||}
\hline
$N$ & $K$ & $|\dgrad|/\milli\tesla\,\meter^{-1}$   & $\tmix/\milli\second$   & $\dlen/\milli\second$   & $\gapone/\milli\second$   & $\dwfactor/\second\,\milli\meter^{-2}$  & $\te/\milli\second$ & $\tr/\milli\second$ \\
\hline
103 & 25 & 300.0 & 6.0 & 5.0 & 0.0 & 2306 & 26.0 & 2600 \\

108 & 25 & 113.5 & 137.0 & 5.0 & 3.4 & 3425 & 26.0 & 2600 \\

78 & 25 & 260.4 & 137.0 & 4.5 & 3.9 & 14631 & 26.0 & 2600 \\
\hline
\end{tabular}
\begin{center}(b)\end{center}
\caption{\label{table:ActiveAxG300}The (a) ActiveAxPGSE and (b) ActiveAxSTEAM protocols. Both come from the experiment design optimisation in~\cite{AlexanderMRM08,DyrbyMRM12} with $\gmax=300\milli\tesla\,\meter^{-1}$. $N$ is the number of diffusion weighted images in each shell. $K$ is the number of nominal $\dwfactor=0$ images associated with each shell.  The total number of images in each protocol is thus $364$. The nominal $\dwfactor=0$ images associated with each shell in ActiveAxSTEAM have the same $\tmix$ as the diffusion weighted images in that shell. The compensated STEAM protocol ActiveAxSTEAMCOMP follows (b), but replaces each $\dgrad$ according to Eq.~\ref{eq:compensation}.}
\end{table}

\begin{table}

\begin{tabular}{||l||l|l|l|l|l|l||l|l|l|l||}
\hline
 & \multicolumn{6}{|c||}{Uncompensated} & \multicolumn{4}{|c||}{Compensated} \\
\hline
 & \multicolumn{2}{|c|}{A1} & \multicolumn{2}{|c|}{A2} & \multicolumn{2}{|c||}{A3} & \multicolumn{2}{|c|}{A1/A2} & \multicolumn{2}{|c||}{A3} \\
\hline
 & $\perp$ & $\parallel$ & $\perp$ & $\parallel$ & $\perp$ & $\parallel$ & $\perp$ & $\parallel$ & $\perp$ & $\parallel$ \\
\hline
FA & $0.513$ & $0.884$ & $0.572$ & $0.495$ & $0.572$ & $0.495$ & $0.576$ & $0.574$ & $0.576$ & $0.574$ \\
std & $0.035$ & $0.171$ & $0.026$ & $0.043$ & $0.026$ & $0.043$ & $0.020$ & $0.021$ & $0.020$ & $0.021$ \\
\hline
$\lambda_1$ & $5.508$ & $1.524$ & $5.270$ & $3.378$ & $5.270$ & $3.378$ & $5.568$ & $5.544$ & $5.568$ & $5.544$ \\
std & $0.260$ & $0.272$ & $0.209$ & $0.187$ & $0.209$ & $0.187$ & $0.161$ & $0.166$ & $0.161$ & $0.166$ \\
\hline
$\alpha$ & $4.603$ & $63.752$ & $2.505$ & $5.085$ & $2.505$ & $5.085$ & $1.921$ & $1.999$ & $1.921$ & $1.999$ \\
$\dirconc$ & $5.351$ & $1.890$ & $6.264$ & $5.345$ & $6.264$ & $5.345$ & $6.791$ & $6.712$ & $6.791$ & $6.712$ \\
\hline
\end{tabular}

\caption{\label{table:AnisoG300} Statistics from simulations with anisotropic DTs for each approximation using the $b=3425\,\second\,\milli\meter^{-2}$ shell of ActiveAxSTEAM, and SNR of 20. The units of $\lambda_1$ are $10^{-10}\,\meter^2\second^{-1}$; the units of $\alpha$ are degrees.  Rows labelled ``std'' show the standard deviation of the quantity above. The true FA is 0.603 and the true $\lambda_1$ is $6\times10^{-10}\,\meter^2\second^{-1}$. Higher $\dirconc$ is better in this experiment.}
\end{table}

\begin{table}

\begin{tabular}{||l||l|l|l||l|l||}
\hline
 & \multicolumn{3}{|c||}{Uncompensated} & \multicolumn{2}{|c||}{Compensated} \\
\hline
 & A1 & A2 & A3 & A1/A2 & A3 \\
\hline
\hline
FA & $0.284$ & $0.175$ & $0.175$ & $0.058$ & $0.058$ \\
std & $0.075$ & $0.033$ & $0.033$ & $0.019$ & $0.019$ \\
\hline
$\lambda_1$ & $3.809$ & $3.660$ & $3.660$ & $4.021$ & $4.021$ \\
std & $0.297$ & $0.168$ & $0.168$ & $0.102$ & $0.102$ \\
\hline
$\dirconc$ & $1.516$ & $0.848$ & $0.848$ & $0.412$ & $0.412$ \\
\hline
\end{tabular}

\caption{\label{table:IsoG300} Statistics, as in table~\ref{table:AnisoG300}, from simulations with isotropic DTs.  The true FA is 0; the true $\lambda_1$ is $4\times10^{-10}\,\meter^2\second^{-1}$. Here $\dirconc$ should be zero.}
\end{table}

\begin{table}

\begin{tabular}{||l||l|l|l|l|l|l||l|l|l|l||}
\hline
 & \multicolumn{6}{|c||}{Uncompensated} & \multicolumn{4}{|c||}{Compensated} \\
\hline
 & \multicolumn{2}{|c|}{A1} & \multicolumn{2}{|c|}{A2} & \multicolumn{2}{|c||}{A3} & \multicolumn{2}{|c|}{A1/A2} & \multicolumn{2}{|c||}{A3} \\
\hline
 & $\perp$ & $\parallel$ & $\perp$ & $\parallel$ & $\perp$ & $\parallel$ & $\perp$ & $\parallel$ & $\perp$ & $\parallel$ \\
\hline
FA & $0.873$ & $0.862$ & $0.862$ & $0.863$ & $0.862$ & $0.863$ & $0.864$ & $0.864$ & $0.864$ & $0.864$ \\
std & $0.018$ & $0.017$ & $0.018$ & $0.017$ & $0.018$ & $0.017$ & $0.017$ & $0.017$ & $0.017$ & $0.017$ \\
\hline
$\lambda_1$ & $15.980$ & $17.508$ & $16.197$ & $16.290$ & $16.197$ & $16.290$ & $16.341$ & $16.344$ & $16.341$ & $16.344$ \\
std & $0.564$ & $0.573$ & $0.573$ & $0.529$ & $0.573$ & $0.529$ & $0.526$ & $0.528$ & $0.526$ & $0.528$ \\
\hline
$\alpha$ & $12.474$ & $2.555$ & $1.450$ & $1.463$ & $1.450$ & $1.463$ & $1.432$ & $1.433$ & $1.432$ & $1.433$ \\
$\dirconc$ & $7.249$ & $7.422$ & $7.359$ & $7.340$ & $7.359$ & $7.340$ & $7.378$ & $7.378$ & $7.378$ & $7.378$ \\
\hline
\end{tabular}

\caption{\label{table:AnisoHuman}Simulation statistics for anisotropic diffusion with the human protocol.  The true FA is 0.87 and the true $\lambda_1$ is $1.7\times10^{-9}\,\meter^2\second^{-1}$.}

\end{table}

\begin{table}

\begin{tabular}{||l||l|l|l||l|l||}
\hline
 & \multicolumn{3}{|c||}{Uncompensated} & \multicolumn{2}{|c||}{Compensated} \\
\hline
 & A1 & A2 & A3 & A1/A2 & A3 \\
\hline
\hline
FA & $0.240$ & $0.099$ & $0.099$ & $0.099$ & $0.099$ \\
std & $0.044$ & $0.032$ & $0.032$ & $0.032$ & $0.032$ \\
\hline
$\lambda_1$ & $5.139$ & $4.324$ & $4.324$ & $4.326$ & $4.326$ \\
std & $0.313$ & $0.252$ & $0.252$ & $0.251$ & $0.251$ \\
\hline
$\dirconc$ & $3.149$ & $0.444$ & $0.444$ & $0.416$ & $0.416$ \\
\hline
\end{tabular}

\caption{\label{table:IsoHuman}Simulation statistics for isotropic diffusion with the human protocol.  The true FA is 0 and the true $E_1$ is $0.7\times10^{-9}\,\meter^2\second^{-1}$.}
\end{table}

\clearpage

\end{document}